\documentclass[conference]{IEEEtran}
\usepackage[utf8]{inputenc}
\usepackage{amsmath,amsfonts}
\usepackage{graphicx}
\usepackage{cite}
\usepackage{siunitx}
\usepackage{hyperref}
\usepackage{float}

\title{Interpretive modeling of plasma evolution during fueling experiments at CMFX}

\author{
\IEEEauthorblockN{
S. Mackie\IEEEauthorrefmark{1}, 
J. G. van de Lindt\IEEEauthorrefmark{1}, 
J. L. Ball\IEEEauthorrefmark{1}, 
A. Perevalov\IEEEauthorrefmark{2}, 
W. Morrissey\IEEEauthorrefmark{3}, 
Z. Short\IEEEauthorrefmark{3},\\
B. L. Beaudoin\IEEEauthorrefmark{3}, 
C. A. Romero-Talamás\IEEEauthorrefmark{2}, 
J. Rice\IEEEauthorrefmark{1},
R. A. Tinguely\IEEEauthorrefmark{1} 
}
\IEEEauthorblockA{\IEEEauthorrefmark{1}MIT Plasma Science and Fusion Center, Cambridge, MA, USA}
\IEEEauthorblockA{\IEEEauthorrefmark{2}University of Maryland Baltimore County, Baltimore, MD, USA}
\IEEEauthorblockA{\IEEEauthorrefmark{3}University of Maryland College Park, College Park, MD, USA}
}

\begin{document}

\maketitle

\begin{abstract}
The Centrifugal Mirror Fusion Experiment (CMFX) is an axisymmetric magnetic mirror with a central cathode which generates an azimuthal, radially sheared, supersonic \( E \times B \) flow. The induced rotation stabilizes, confines, and heats the plasma. The diagnostic set on CMFX is sparse, giving limited insight to the state of the plasma. In this work, we developed a time-dependent interpretive analysis framework that uses applied voltage, input power, and measured neutron yield rate to infer evolving plasma conditions throughout a discharge. The 0D MCTrans++ code serves as the core physics model, incorporating centrifugal effects, viscous heating, and angular momentum confinement to infer plasma parameters from operating conditions and experimental observables. An iterative Newton's method was implemented to solve for the plasma state evolution consistent with experimental measurements averaged over successive time intervals. The interpretive analysis was applied to experiments comparing different fueling strategies, revealing a path to improved performance via several short puffs of fuel spread across the discharge. This insight led to operations at voltages up 70 kV. Deuterium neutron yields up to \(1.5 \times 10^7\) n/s were measured, and ion temperature was inferred to reach 950 eV. Until CMFX gains a more complete diagnostic set, this interpretive analysis framework provides useful insight into the evolution of centrifugal mirror plasmas.
\end{abstract}

\section{Introduction}
\begin{figure}[H]
    \centering
    \includegraphics[width=\linewidth]{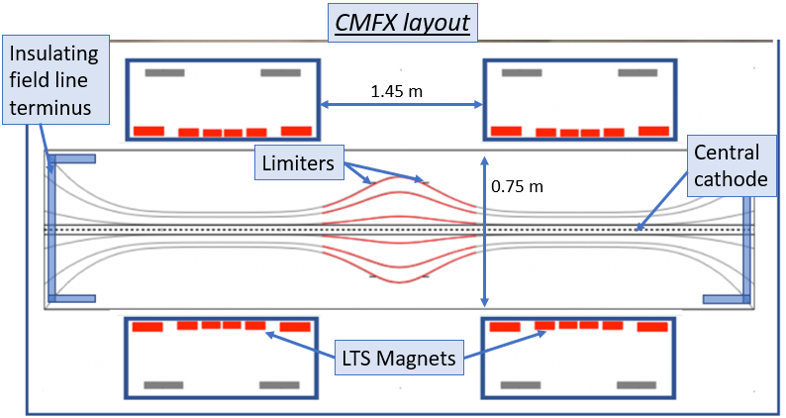}
    \caption{Schematic layout of CMFX showing the major components of the machine and the magnetic field geometry (not to scale). Two large-bore LTS magnets produce the mirror field. The electric field is generated by the central cathode and tungsten-coated ring electrodes, which also serve as plasma limiters. Insulating end plates prevent electrical shorting of the radial electric field along magnetic field lines. Red field lines indicate where the field was measured to validate the calculated field model.}
    \label{fig:schematic}
\end{figure}

The Centrifugal Mirror Fusion Experiment (CMFX) is a rotating mirror fusion device designed to explore operational regimes toward those relevant to future power reactors.  CMFX builds upon the results of the Maryland Centrifugal Experiment (MCX); namely, that a sheared, supersonic \( E \times B \) flow driven by a strong negatively biased central electrode stabilizes MHD instabilities, suppresses turbulent transport, and viscously heats the plasma\cite{Ellis2001}\cite{Ellis2005}\cite{Romero-Talamás2012}.  Figure \ref{fig:schematic} shows a schematic of CMFX's main structures as well as the vacuum magnetic flux surfaces. CMFX plasmas are confined by a pair of Low Temperature Superconducting (LTS) magnets, a negative high voltage central cathode, and a vacuum chamber. The LTS magnets are separated by 145 cm and produce a 3 T magnetic field at the mirror throat and 0.3 T at the midplane giving a mirror ratio of 10. The vacuum chamber is a cylinder 75 cm in diameter. A high voltage DC power supply biases the central cathode to up to 100 kV and can deliver up to 1 A, although a protection circuit automatically triggers to reduce the voltage when the current reaches 800 mA. The plasma shape is limited by two annular tungsten-coated grounding electrodes with a 44 cm diameter. Further description of the CMFX machine may be found in \cite{Ball2025}.

Recent work has focused on characterizing the fusion neutron yield from deuterium discharges on CMFX\cite{Ball2025}. The neutron yield rate has been measured with two independently calibrated diagnostics: a $^3$He gas-filled proportional counter, and a set of liquid organic scintillator detectors\cite{Ball2024}. The $^3$He counter was calibrated \textit{in situ} using a NIST-traceable $^{252}$Cf source, while the scintillators were calibrated \textit{in silico} using coupled OpenMC and GEANT4 simulations of neutron transport and detector response \cite{ROMANO2015} \cite{geant4}. These two methods were shown to agree very well within uncertainties; a linear fit comparison of the two inferred yields gave a slope of 0.98 with an $R^2$ coefficient of 0.978, giving confidence in the accuracy of the measured neutron yield rate.  It was observed that fusion yield was extremely repeatable for fixed discharge parameters and that the yield increases exponentially with applied voltage up to about 65 kV. Higher voltages were inaccessible during previous experiments due to arcing terminating plasma operations.

An interpretive 0D physics model was developed to infer discharge-averaged plasma parameters such as ion temperature, density, and confinement times from measurements of the bias voltage, input power, and neutron production \cite{Ball2025}. The model is built around MCTrans++, a 0D centrifugal mirror plasma code developed to solve conservation equations for particle density, energy, and angular momentum in the limit of strong magnetization, low-collisionality, supersonic azimuthal flow, large mirror ratio, and large applied bias compared to the ambipolar potential. The code has been benchmarked against experimental data from previous rotating mirror experiments\cite{mctrans}. Transport and confinement parallel to field lines is derived following Pastukov's work on electrostatic confinement of electrons in mirrors, modified to account for the additional centrifugal potential \cite{Pastukhov_1974}. Radial transport across flux surfaces is assumed to be classical due to flow-shear suppression of turbulence \cite{Huang2001}.

In a standard MCTrans++ run, the electron density, neutral density, bias voltage, device and plasma geometries, magnetic field strength, and effective charge state are supplied as inputs; and the steady state volume average temperature, confinement times, particle loss rates, fusion rates, required input power, etc., are calculated accordingly. However, in our interpretive workflow, an iterative Newton's method is implemented to invert the model and find the electron and neutral densities which reproduce the experimental measurements. After specifying the magnetic field based on measurements and fixing our assumed plasma length at 60 cm and effective charge state at 3, the model has three independent unknown variables: ion temperature, electron density, and neutral density. Thus, using three measured variables, we can invert the system. Once MCTrans++'s calculated input power, fusion rate, and bias voltage converge to the observed values, the computed values of the other plasma parameters are taken as inferences of the true experimental conditions. 

Applying this workflow to a scan in voltage revealed that the ion temperature scales linearly with applied voltage, $T[\mathrm{keV}] = 0.014V[\mathrm{kV}]-0.174$ ($R^2=0.996$), while energy confinement time is independent of voltage, a conclusion consistent with viscous heating proportional to applied voltage \cite{Ball2025}. The electron density was found to not depend strongly on the applied voltage, but did show variation with the amount of fuel initially injected.

This article extends the interpretive physics model to extract the time evolution of the 0D plasma state parameters. This time-dependent interpretive modeling framework is described in detail in section \ref{sec-interp}. We apply this model to two sets of experiments in section \ref{sec-results}: first, a set of discharges designed to compare the effects of different fueling schemes; second, a set of 30 repeated discharges with an optimized fueling scheme that achieves high plasma performance. We discuss the significance and limitations of these results in section \ref{sec-disc}. We conclude in section \ref{sec-conc} by summarizing our findings and identifying key areas of future work.

\section{Time-dependent Interpretive Modeling}\label{sec-interp}

MCTrans++ \cite{mctrans} implements a set of multi-fluid conservation equations for each plasma species' population

\begin{equation}
    \frac{\partial N_s}{\partial t} + \frac{1}{R} \frac{\partial}{\partial R}\left(R\Gamma_s\right)=S_{n,s},
\end{equation}\label{eq:ParticleConservation}

stored energy associated with each plasma species

\begin{equation}
    \frac{\partial}{\partial t}\left(\frac{3}{2}N_s T_s  \right) + \frac{1}{R}\frac{\partial}{\partial R}\left(R q_s \right)=\pi_s^{(R\phi)}\frac{\partial \omega}{\partial R}+Q_s+S_{E,s},
\end{equation}\label{eq:StoredEnergy}

and the angular momentum of the plasma
\begin{equation}
    \frac{\partial}{\partial t}(J\omega)+\frac{1}{R}\frac{\partial}{\partial R}\left(R\sum_s \pi_s^{(R\phi)}\right)=-j_R RB + S_\omega .
\end{equation}\label{eq:AngularMomentum}

Here, the subscript \textit{s} is a label for each species, and $R$ is the radial coordinate. $N_s$ is the midplane particle density, $T_s$ is the temperature, and $\omega$ is the azimuthal angular velocity of the plasma. $\Gamma_s$, $q_s$ and $\pi_s^{R\phi}$ are the radial particle, heat, and azimuthal angular momentum fluxes, respectively; and $S_{n,s}$, $S_{E,s}$ and $S_\omega$ are arbitrary sources of particles, energy and angular momentum. $J$ is the moment of inertia of a given flux surface. $j_R$ is the radial current density driven by the external high voltage power supply, which spins the plasma up and drives the viscous heating represented by the term $Q_s$.

A series of simplifying assumptions are made which reduce this set of differential equations to an algebraic system. For more details about the computational physics model implemented by MCTrans++ and benchmarks against experiments, see Schwartz 2024\cite{mctrans}. Given data on machine geometry, applied electric and magnetic fields, and plasma composition, MCTrans++ calculates equilibrium 0D values of plasma parameters such as the average ion and electron temperature, plasma and neutral densities at the machine midplane, total charge exchange loss rate, total fusion rate, input power, and confinement times. 

In \cite{Ball2025}, we developed a Newton's method solver that, given an experimentally applied voltage, iterates the electron and neutral density until the model's prediction for the fusion rate and required input power converge with the experimentally observed values. We then take the model's predictions of the plasma equilibrium state to represent the experimental state of the plasma. No attempt is made to propagate measurement uncertainties through the solver. Nonetheless, this method allowed us to study qualitative trends in shot-averaged values as a function of applied voltage. For detailed discussion about the specific assumptions about CMFX machine parameters used in our modeling and model uncertainties, see \cite{Ball2025}.

In this work, we extend the interpretive analysis to study the time evolution of the equilibrium of CMFX. By subdividing the voltage flattop phase of the discharge and applying the interpretive analysis previously developed to each subinterval, we estimate the evolution of the plasma's state. Each subinterval is 25 ms long. For each subinterval, the average neutron rate, bias voltage, and current draw are calculated from experimentally measured data.  These data are fed into the 0D interpretive physics model to infer the average densities, temperatures, and other plasma parameters. The converged values of the model were used as the initial guess for the next time step's solver to improve computational efficiency. Solving the inversion of MCTrans++'s physics model from our measurements requires a neutron rate greater than approximately $10^6$ neutrons per second, which sets the start and end of inferred data series. This puts the full ramp up and termination phases of the discharges out of range of this analysis.


25 ms is much longer than typical timescales for plasma equilibration. The thermal transit time is $\tau_{th,i}=L/v_{th,i}=3.8\mathrm{\mu s}$, assuming a 500 eV ion temperature and 60 cm plasma length. The MHD timescale is even faster; taking a typical ion number density of $10^{18}\mathrm{m^{-3}}$ and field strength of 300 mT gives $\tau_{MHD}=L/v_A=0.1 \mathrm{\mu s}$. Even during transient phases, such as the ramp up and ramp down of the plasma or mid-discharge gas puff fueling (see Fig. \ref{fig:fueling} for example), where time dependent dynamics may become significant, we expect the plasma to evolve through states not far from thermal equilibrium, and we interpret the neutron yield rate observed over that interval as representing the average thermal reactivity as the transient decays to steady state. This analysis breaks down if the fusion is not thermonuclear; however, previous centrifugal mirrors have been shown to be in local thermodynamic equilibrium \cite{Ellis2005}, so we expect CMFX is as well. This assumption should be explicitly confirmed with neutron anisotropy measurements in the future.

\section{Results}\label{sec-results}

We apply the time resolved interpretive analysis to experiments designed to investigate the impact of fueling on plasma performance. In these discharges, CMFX was fueled with pure deuterium by a programmable gas valve located at the machine midplane. In all cases, the fuel lines are pre-filled with deuterium gas at fixed pressure and the valve is opened at the programmed times for the specified duration. The vacuum vessel is pumped by two non-evaporative getter (NEG) pumps (one at each end of the mirror, near the insulators, see Figure \ref{fig:schematic}) and a turbomolecular pump placed far from the magnets.

The amount of deuterium injected into the machine can estimated by assuming the flow through the nozzle is choked and therefore the mass flow rate is given by:

\begin{equation}
    \dot m = C_d AP_0\sqrt{\frac{\gamma}{RT}\left(\frac{2}{\gamma+1} \right)^\frac{\gamma+1}{\gamma-1}}
\end{equation}

Here $\dot m$ is the mass flow rate; $C_d$ is the discharge coefficient, a dimensionless parameter representing departure from ideal choked flow and taken here to be 0.5; $A$ is the area of the nozzle, which has a 0.1016 mm diameter; $P_0$ is the back pressure in the gas line, held constant at 80 psi; $\gamma$ is the ratio of heat capacities and is 1.4 for a diatomic gas; $R$ is the ideal gas constant; and $T$ is the absolute temperature of the gas, taken to be room temperature. The total amount of deuterium injected is then estimated by multiplying by the duration that the valve is opened.

Uncertainty is quantified for the measured values of bias voltage, current draw, input power and neutron emission rate. However, the uncertainty in the power supply readings was found to be negligible and so is not included in the plots. Neutron error bars include Poisson statistics and calibration uncertainty. As explained above and in \cite{Ball2025}, the model uncertainty and bias are difficult to quantify and so the inferred values are presented without error bars.

\subsection{Study on gas puff fueling at 55 kV}
To study the effects of fueling on fusion performance, discharges at 55 kV were analyzed with three different fueling schemes and otherwise identical conditions:

\begin{itemize}
    \item Shot 1995: Single 1 ms gas puff (0.48 $\mu\mathrm{Mol}$ $\mathrm{D_2}$),
    \item Shot 2042: Single 0.25 ms gas puff (0.12 $\mu\mathrm{Mol}$ $\mathrm{D_2}$),
    \item Shot 2031: Two 0.25 ms gas puffs spaced by 400 ms.
\end{itemize}

\begin{figure}
    \centering
    \includegraphics[width=\linewidth]{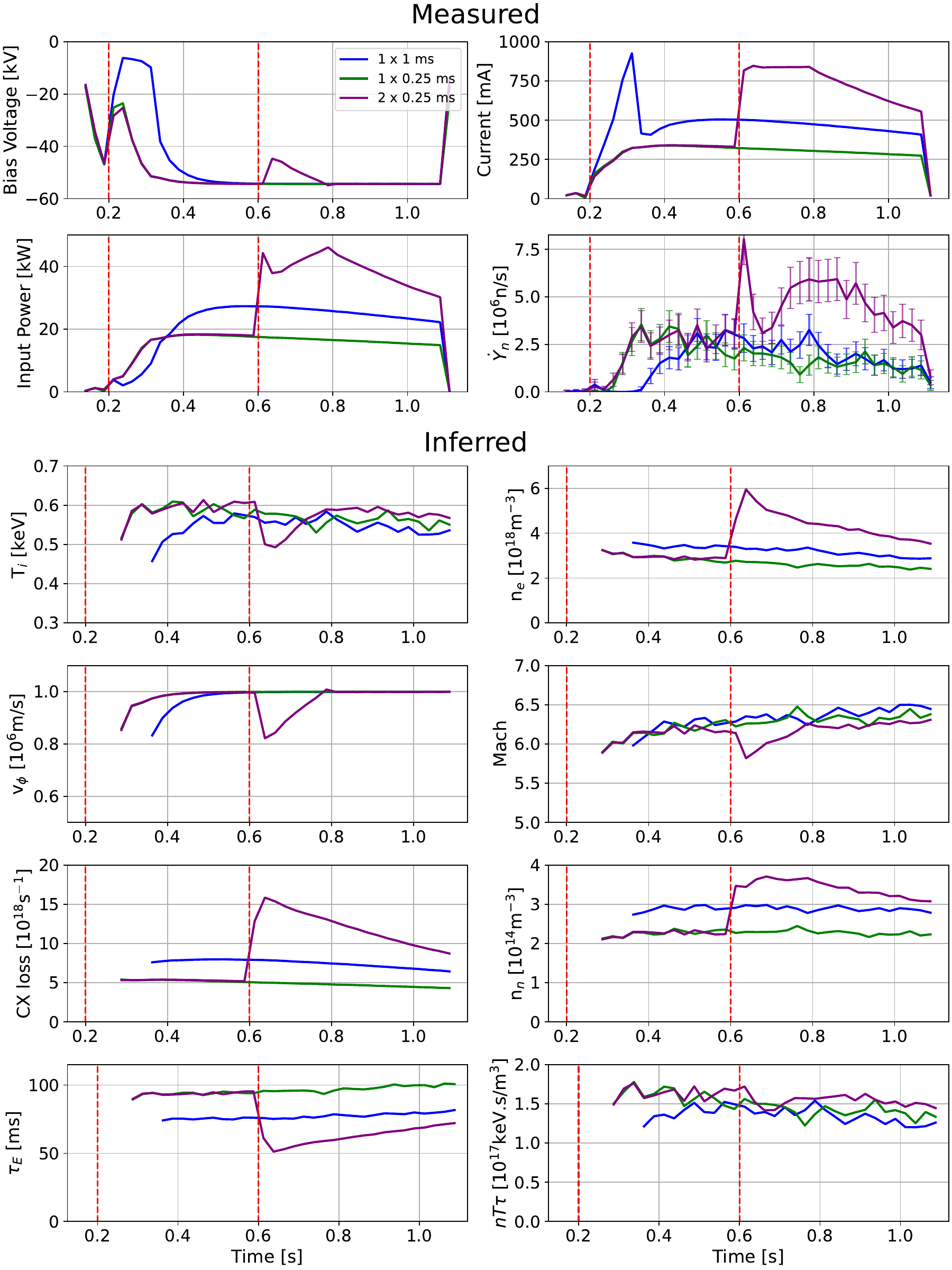}
    \caption{Measured and inferred evolution of CMFX plasma under different fueling schemes - blue, single 1 ms fuel puff; green, single 0.25 ms fuel puff, purple, double 0.25 ms fuel puff. The red dashed vertical lines indicate when the  fuel was injected into the machine. Top two rows show measured applied bias voltage, current, input power and neutron yield rate, while remaining plots show plasma parameters inferred by MCTrans++ interpretive model: ion temperature, electron density, azimuthal velocity, mach number, charge exchange loss rate, neutral density, energy confinement time, and triple product.}
    \label{fig:fueling}
\end{figure}

Figure \ref{fig:fueling} shows the measured traces of bias voltage, current, input power and neutron yield rate in the first two rows and the inferred evolution of the plasma state in the remaining panels for all three experimental conditions. The neutron yield rate reported for these experiments was measured using the liquid scintillator described above and more fully in \cite{Ball2025}. 

The measured traces of the short puff discharges are indistinguishable until the second puff is injected, speaking to the high repeatability of CMFX discharges. The long puff takes about 200 ms longer to ramp up to flat top conditions but ultimately achieves a yield rate comparable to the single short puff. When additional fuel is injected, the plasma tries to draw more current to ionize the additional gas until the power supply's protection circuit triggers, throttling the current at 800 mA by reducing the bias voltage until the current drops back to the supply's operating range. This causes the neutron yield rate to dramatically spike before rising to triple the level observed during single puff discharges once the voltage stabilizes at the set point again. 

The inferred ion temperature traces are qualitatively similar in all cases reaching 550-600 eV. The end of the initial heating during ramp up is captured in the first few time intervals of the analysis. The long puff discharge heats more slowly than the short puff cases. This is because more neutral gas must be ionized and spun up to stabilize the configuration which takes more time. The second injection of fuel initially cools the plasma and although it does recover, the heating is more gradual than the initial ramp-up transient appears.

No initial transient in the electron density traces is resolved by our analysis. In all cases, the density decays slowly from about 3$\times 10^{18}$ $\mathrm{m}^{-3}$. The long puff discharge has a slightly higher density. Injection of additional fuel causes the electron density to double, but this elevated density decays more rapidly than before additional fuel was injected. This is because the injection of gas increases the neutral inventory which acts as a cold stationary background to the supersonic rotating plasma. Scattering and charge exchange increase the loss rate of plasma particles from the confining potential.

The rotational velocity traces indicate that the plasma is rapidly torqued by the electromagnetic field to a steady and consistent rotational velocity of 1,000 km/s. The fact that all conditions achieve the same equilibrium speed reflects the independence of $E\times B$ motion from the plasma density. The drop in rotation speed following the mid-discharge fuel injection is due to the drop in applied voltage required to accommodate the machine protection current throttle. The Mach numbers' slow growth indicates the sound speed is decreasing as a result of slow decay of the thermal pressure (both density and temperature slowly decay over the course of the discharge).

Charge exchange and neutral density are fairly constant over the discharge in all cases, except when additional fuel is injected after which both rapidly rise. The 1 ms fuel puff discharge is inferred to have about 40\% higher charge exchange loss rate and neutral density as a result of more gas being initially injected.

The energy confinement time is high, approaching 100 ms in the short puff discharges, and improves over the course of the discharges. The injection of fuel reduces the energy confinement time by a factor of two. The long puff confinement time is 80\% that of the short puff discharges.  The triple product is nearly identical in all three cases, showing a slight decline in time. The similarity in triple product traces indicates that shorter fuel puffs perform similarly to or better than longer fuel puffs. Injecting additional fuel during a discharge opens access to higher density and does not significantly degrading the plasma's performance, although care must be taken to stay within the operating limits of the power supply. 

\subsection{High Performance Operation at 70 kV}

\begin{figure}
    \centering
    \includegraphics[width=\linewidth]{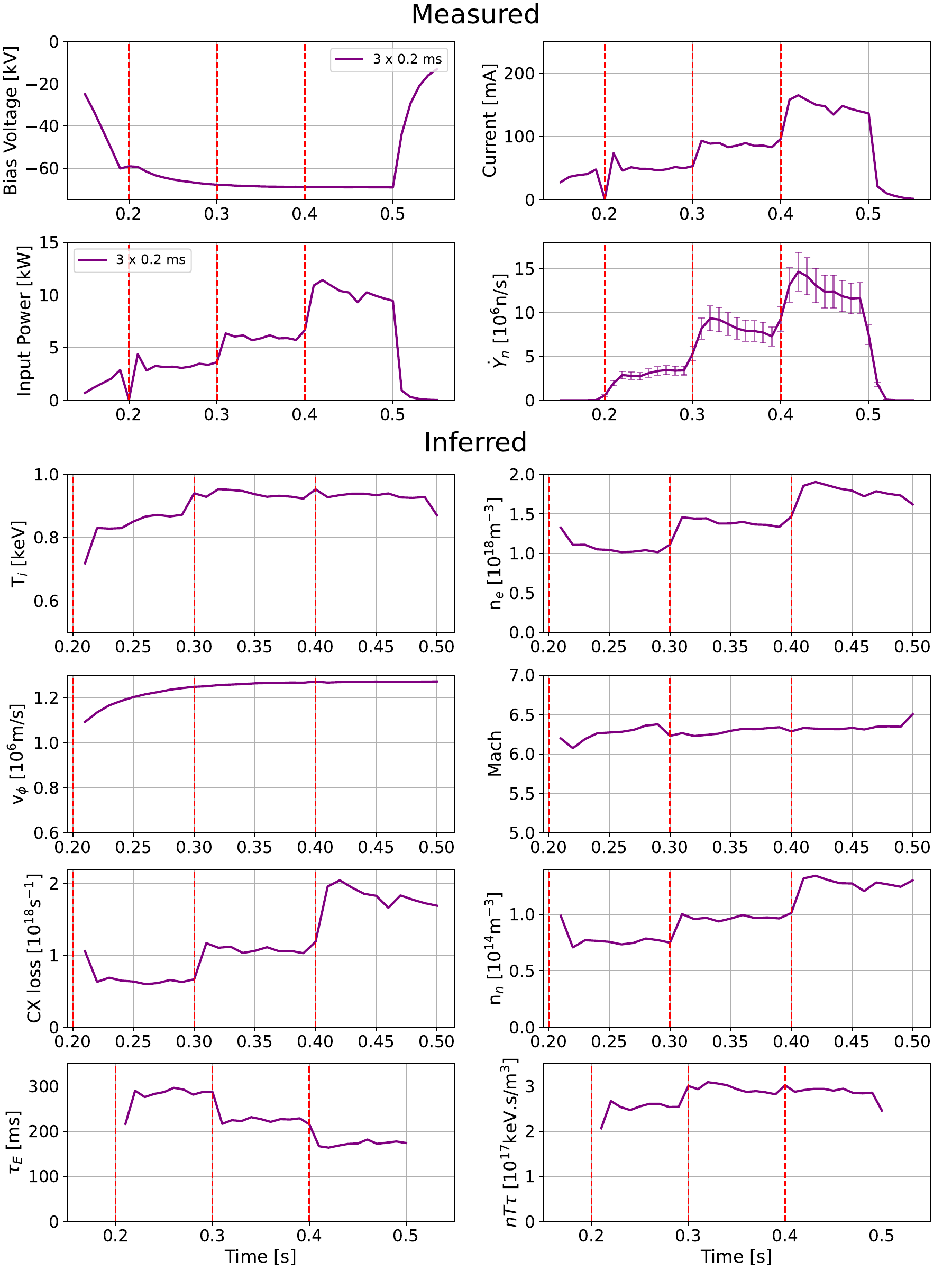}
    \caption{Time traces of measured and inferred plasma parameters at 70 kV. The first two panels show the average traces of applied bias voltage, current draw, input power and neutron yield measured for 30 repeated discharges. The remaining panels show the results of our interpretive physics modeling: ion temperature, electron density, azimuthal velocity, mach number, charge exchange loss rate, neutral density, energy confinement time, and triple product. Red vertical lines indicate when gas is injected into machine.}
    \label{fig:high}
\end{figure}

With the insights gained from the fueling study, operation at even higher voltages was attained by further reducing the amount of gas injected in each puff. A series of 30 identical discharges was repeated at 70 kV. Each discharge was fueled by three 0.2 ms deuterium gas puffs spaced by 100 ms. Figure \ref{fig:high} shows the measured traces of voltage, current, power and neutron rate averaged over the 30 replicate experiments as well as the inferred average plasma evolution. The neutron rate was measured with the He-3 proportional counter for these discharges.

The measured power draw and neutron yield both clearly exhibit stepwise increases in response to fuel injection. Notably, the power draw is significantly reduced compared to the previous discharges shown in Figure \ref{fig:fueling}, while the measured neutron rate is initially similar. Comparing the initial power draw in the 0.25 ms puff experiments, reducing the fuel puff duration by 20\% yields an 80\% reduction in the power draw despite the higher operating voltage. This is due to the reduced current draw, indicating a lower plasma resistivity. It also reflects the fact that less energy is required to ionize and torque the plasma when less fuel is injected. Furthermore, the power supply does not enter a current-limited mode when additional gas is injected and no transient spike is observed in the neutron yield rate.  A peak neutron rate of 1.5$\times10^7$ n/s is observed following the third gas injection, the highest yield rate reported thus far on CMFX.

Interpretive modeling gives insight to the evolution of the plasma state under these conditions. The ion temperature initially rises to approximately 850 eV during the initial ramp up phase, matching the linear trend observed in \cite{Ball2025} well. The second injection of fuel causes the temperature to rise to 950~eV, while the third puff does not trigger further heating of the plasma. Explaining the anomalous increase in temperature with the second puff but not third will be a topic of future work with more sophisticated diagnostics.

The electron density is inferred to be about $10^{18}$ $\mathrm{m}^{-3}$ after the plasma is initiated, and increases by about $5\times10^{17}$ $\mathrm{m}^{-3}$ with each additional gas puff. Following each injection, the decay rate of the plasma density is slightly faster, consistent with the higher inferred charge exchange loss rate. The density is about 1/3 the level inferred in the lower voltage 0.25 ms fuel puff experiments in Figure \ref{fig:fueling}.

The plasma rotation spins up smoothly to approximately 1,250 km/s in these discharges. Unlike in Figure \ref{fig:fueling}, the rotation is not slowed by the injection of additional fuel. This is because the 20\% reduction in the fuel injection avoids overburdening the power supply. The applied voltage is not reduced by the machine protection circuit to throttle the current, so the radial electric field is constant as the fuel is injected. The Mach number, however, does evolve non-monotonically over the course of the discharge as a result of the plasma pressure responding to the fuel injections.

The charge exchange loss rate is initially about 10 times lower than in Figure \ref{fig:fueling}. This indicates that the charge exchange is very sensitive to neutral inventory and fueling parameters. The neutral density is about 3 times lower initially. Both the neutral density and charge exchange loss rate increase sharply in response to the gas puffs. The inferred energy confinement time starts at about 300 ms and decreases with each fuel puff, but even after the third puff is still nearly double the highest value observed in the fueling study at 55 kV. The triple product trace qualitatively appears similar to the temperature trace, as the increase in density is balanced by reduction of the confinement time. After the second puff, the triple product is fairly constant at $3\times10^{17}$keV s m$^{-3}$, double the value in the 55 kV experiments.

\section{Discussion}\label{sec-disc}

These results suggest interesting trends in the evolution of CMFX plasmas under different fueling schemes that would be inaccessible from raw data alone, and have opened several avenues of investigation for improving fusion performance. The most important takeaway is that machine operation is sensitive to fueling, while the plasma's fusion performance is less so. Too much gas causes arc discharges from the central cathode, prematurely terminating the discharge and potentially damaging machine components. Comparing the green and blue traces in Figure \ref{fig:fueling} demonstrated that reducing the initial fuel injection avoids this operational limit, opening up higher voltage regimes while maintaining or improving plasma performance. 

Furthermore, gas puff fueling mid-discharge has been demonstrated as an effective way to increase the density of CMFX plasmas and thereby boost fusion rates. The addition of gas increases the load on the power supply as the gas is ionized and spun up, and it is important to keep the additional current required to maintain the bias voltage below 800 mA to avoid triggering the power supply's protection circuit. This operational current limit was avoided by slightly reducing the gas injection duration from 0.25 ms to 0.2 ms; this allowed the voltage to remain constant as more fuel was added which caused an increase rather than a decrease in temperature (compare Figures \ref{fig:fueling} and \ref{fig:high}).

These fueling insights enabled relatively high plasma performance using a three-puff fueling scheme at 70 kV. This discharge was repeated 31 times and was prematurely terminated only once by an arc, showing that this is a robust path toward higher voltage and fusion performance. In these discharges, we infer an energy confinement time of 200-300 ms, rotational velocity 1,250 km/s, ion temperatures approaching 1 keV, and a fusion triple product of 3$\times 10^{17} \mathrm{keV s/m}^3$. These performance metrics, if corroborated by direct measurements, would place CMFX among the highest performing magnetic mirror experiments in the literature to date, second only to the GOL-3 mirror \cite{tripleproduct}. Compared to tokamaks, these results place CMFX between ST and low performance discharges at EAST. Future hardware upgrades are expected to significantly increase the performance of CMFX. 

While MCTrans++ has been benchmarked against experimental data from previous machines such as IXION, it has not yet been rigorously benchmarked against CMFX experiments \cite{mctrans}. Realistic assumptions were made about input parameters, such as effective ion charge $Z_{eff}$ and plasma length, that slightly alter the inferred plasma state. The parameters introduce a bias in the interpretive analysis through the plasma volume, resistivity, radiation loss rate, etc., used to solve the system of equations. As such, the interpretive modeling results presented here should not be confused for measurements. These results must be validated by actual measurements as additional diagnostics are installed and commissioned on CMFX. 

In spite of the potential model error, the utility of this interpretive model is clear. Its power lies in the relative simplicity of centrifugal mirrors and the strong coupling between neutron yield rate and important plasma parameters. With a minimal diagnostic set consisting of neutron counters and the read-out of the high voltage DC power supply, we were able to infer the evolution of key reactor performance metrics including ion temperature, plasma density, energy confinement time, and rotation speed.  If direct measurements of the plasma temperature and density agree with those inferred by this model, it will be a demonstration of the ability of neutron diagnostic systems to provide critical reactor state information using reduced physics models. 

\section{Conclusion and Future Work}\label{sec-conc}
This study presented an implementation of a time-resolved interpretive physics model for analyzing CMFX discharges. Measurements of fusion yield and power supply readings were fed to a 0D physics model to infer the plasma state by iteratively calling MCTrans++ to identify plasma and neutral densities consistent with those measurements. The analysis uses a minimal diagnostic set to extract rich information about the plasma's evolution.

This analysis enabled greater understanding of the role of fueling parameters in determining plasma performance. It was found that reducing the amount of gas puffed into the chamber to approximately 100 nMol improves confinement by reducing charge exchange with neutrals, while overall fusion performance is maintained. In addition, it was found that gas puff fueling mid-discharge is an effective strategy for increasing the plasma density in high voltage operation, although care must be taken to ensure that the power supply does not become current-limited as the additional gas is ionized and spun up. These insights revealed a path to higher voltage operation by making several short gas puffs to avoiding disruptive arcs triggered by neutral gas accumulation.

The absolute performance inferred during the 70 kV advanced fueling discharges is also noteworthy. Achieving ion temperatures on the order of 1 keV is an important milestone, and appears to be within reach for CMFX with slightly higher operating voltages. Extrapolation of the trend from \cite{Ball2025} predicts 1 keV ion temperatures at 84 kV bias for single puff discharges. The long energy confinement time is also encouraging for reactor scaling. The low density of CMFX discharges is a challenge that must be addressed in order to scale to power reactor relevant regimes, but these initial investigations into advanced fueling schemes indicate a path to higher densities while avoiding arcing limits.

The experimental results presented in this article demonstrate that CMFX produces highly reproducible, warm, stable plasmas. Future efforts will focus on installing and commissioning additional diagnostics to corroborate our interpretive model. The planned diagnostic upgrades include installing a Thomson scattering system to measure the electron temperature, and implementation of advanced analysis routines for the installed two-color interferometer to deconvolve the effect of radiation heating of the ZnSe vacuum window to extract plasma density. Experiments will also be conducted to continue pushing the operating voltage beyond the 70 kV reported here.

\section*{Acknowledgments}
We would like to thank Drs. Ian Abel and Nick Schwartz for valuable discussions about MCTrans++, Professor Zach Hartwig for allowing us use of the 10 inch liquid scintillator, and the EHS staff at UMD for their assistance in this work.  
CMFX is funded by ARPA-E grant DE-AR0001270.  
2 inch liquid scintillators and DAQ hardware used in this work was acquired by MIT PSFC under Commonwealth Fusion Systems RPP 031.

\bibliographystyle{IEEEtran}
\bibliography{refs.bib}

\end{document}